\documentclass[reprint, amsmath, amssymb, aps, prx]{revtex4-2}

\usepackage{graphicx} 
\usepackage{dcolumn}
\usepackage{bm}
\usepackage{blindtext}
\usepackage[utf8]{inputenc}
\usepackage[T1]{fontenc}
\usepackage[a4paper, margin=2.7cm]{geometry}
\usepackage{amsmath, amsthm, amsfonts, amssymb}
\usepackage{mathrsfs}         
\usepackage{bbm} 
\usepackage{natbib}
\usepackage{url}
\usepackage{upgreek}
\usepackage{braket}
\usepackage{graphicx}
\usepackage{tabularx}
\usepackage{verbatim} 
\usepackage[table]{xcolor}
\usepackage{makecell}
\usepackage{hyperref}

\begin{document}

\title{A Novel Quantum Algorithm for Efficient Attractor Search\\ in Gene Regulatory Networks}

\author{Mirko Rossini$^{1,2,*}$, Felix M. Weidner$^{3,*}$, Joachim Ankerhold$^{1,2}$, Hans A. Kestler$^{3,\dagger}$}
\affiliation{$^1$Institute for Complex Quantum Systems, Ulm University, 89069 Ulm, Germany \\
$^2$Center for Integrated Quantum Science and Technology (IQST) Ulm-Stuttgart, Germany \\
$^3$Institute of Medical Systems Biology, Ulm University, 89069 Ulm, Germany}
\altaffiliation{$^*$Equal contribution, $^\dagger$To whom correspondence should be addressed: [hans.kestler@uni-ulm.de]}

\begin{abstract}

The description of gene interactions that constantly occur in the cellular environment is an extremely challenging task due to an immense number of degrees of freedoms and incomplete knowledge about microscopic details. Hence, a coarse-grained and rather powerful modeling of such dynamics is provided by Boolean Networks (BNs). BNs are dynamical systems composed of Boolean agents and a record of their possible interactions over time. Stable states in these systems are called attractors which are closely related to the cellular expression of biological phenotypes. Identifying the full set of attractors is therefore of substantial biological interest. However, for conventional high performance computing this problem is plagued by an exponential growth of the dynamic state space. 
Here we demonstrate a novel quantum search algorithm inspired by Grover's algorithm to be implemented on quantum computing platforms. The algorithm performs an iterative suppression of states belonging to basins of previously discovered attractors from a uniform superposition, thus increasing the amplitudes of states in basins of yet unknown attractors. This approach guarantees that a new attractor state is measured with each iteration of the algorithm, an optimization not currently achieved by any other algorithm in the literature. Tests of its resistance to noise have also shown promising performance on devices from the current Noise Intermediate Scale Quantum Computing (NISQ) era.

\end{abstract}

\date{\today}
\maketitle

\newpage

\section{Introduction}
DNA encodes the basic information for the construction of all cellular life on earth. 
Its information is translated into the synthesis of proteins which serve as tools for a variety of intracellular tasks. One very important one is the regulation of the translation process as such (gene expression).
The immense number of agents and the lack of knowledge about microscopic details seem to render a dynamical description of gene regulation unfeasible. However, in the last decades, driven by advances in technology and computational power, it has turned out that in many cases a coarse-grained effective modelling is sufficient and powerful to examine crucial aspects of entire networks of genes, mRNAs, and proteins 
as well as to provide detailed descriptions of, for example,  signaling cascades and crosstalk between pathways\cite{Siegle2018}. 

In general, networks consisting of genes
and their mutual interactions are called Genes Regulatory Networks (GRNs). These networks aim to capture the most relevant mechanisms for the self-regulation of the cell's metabolism. Advanced numerical methods have been developed to tackle their complexity,  among many, ranging from differential equations\cite{Davidich2008} to methods based on Petri nets\cite{petrinets}, Boolean Networks (BNs) are considered as one of the most powerful ones. They are able to capture essential phenomena observed in experiments, they can easily be adapted to include emerging additional information,  and they allow to make accurate predictions about the behaviour of a system and, i.e.\ more generally, its phenotype\cite{schwab2020concepts}.

BNs have already been investigated for the analysis of a wide range of systems, including development\cite{Giacomantonio2010}, cell cycle regulation\cite{Faur2006}, hematopoietic stem cells\cite{Ikonomi2020}, and cancer\cite{Werle2021,Cohen2015}. In particular, the problem of attractor search in BNs, which we define mathematically later in this manuscript, is a fundamental task in the field of system biology, as these specific states of the system are often related to the genetic expression and phenotype of the cell. However, the problem of identifying all attractor states by exhaustive simulation of the system dynamics requires the computation of the full State Transition Graph (STG), a directional graph consisting of all possible $2^n$ configurations for $n$ genes, and its evolution in time (via directional links), thus making it an NP-hard problem\cite{akutsu1998system, Akutsu2012, Mori2022}. 

However, the benefits of gaining insight into the dynamics and properties of BNs are not limited to the field of systems biology, as such a method can model any system of interacting Boolean agents. One example currently being investigated for its potential applications as a quantum simulation platform is the transverse Ising model, which is a chain of interacting $1/2$ spins known to oscillate between "state up" and "state down". As several experiments on such platforms are currently being explored, including artificial spin ice (ASI) arrays\cite{Bingham2021}, trapped atomic ions (\cite{Kim2011} and therein), neutral atoms\cite{Graham2022} and superconducting circuits\cite{Kim2023}, etc., the use of BNs could instead advance the development of theoretical predictions about the collective behavior of these systems. This includes possible more exotic collective behaviors such as those observed in topological models such as the Kitaev-Heisenberg model\cite{Gohlke2017}.

The fundamental idea of the presented algorithm is to exploit the interference properties of quantum devices (which arise naturally from their wave-matter duality) to increase the efficiency of the attractor search in a BN. Indeed, we prove that it is possible to design such interferences to suppress any basin of attraction of one or more previously found attractors. This ensures that different subsequent runs of this algorithm would identify a different attractor at each different run, achieving the highest level of optimization possible for a classically inspired search method based on multiple simulations and measurements of a state after it has evolved in time.

We prove the efficiency of our method on two model BNs, one created ad hoc and the other taken from the literature\cite{Giacomantonio2010}. By running our algorithm on a quantum computer simulator, we show the exactness of the results for the problem at hand. We then run the algorithm on a quantum computer simulator capable of mimicking the noise profile of a target quantum device (provided by IBM and the Qiskit Python package) to show the effect of noise on the final results. These results suggest that the algorithm provides solid and reliable results even on today's quantum devices (specifically the \textit{ibm\_brisbane} platform), regardless of their NISQ nature. In the last part, before the conclusions, we provide a detailed breakdown of the algorithm, both in its circuitry and in its mathematical formulation.

\section{Results}
A Gene Regulatory Network can always be mapped onto a dynamical network of Boolean variables, i.e. a Boolean network. Characterising the properties of such maps is therefore crucial for uncovering the molecular mechanisms behind the phenotypic expression of a cell, including genetic patterns that lead to catastrophic consequences such as cancerous behaviour or genetic diseases. Because genetic interactions in cells occur on a very fast time scale, not all possible states that the Boolean network can assume are relevant to the phenotypic expression of a cell, but only those that are stable and do not change over time. Such states are called attractors of the Boolean network.

Many different classical methods have been developed to obtain more or less accurate results for the problem of determining the attractors of a Boolean network, taking advantage of the better efficiency of algorithms that require less precision. In this work, we prove an algorithm for quantum computation that, by representing the problem in terms of qubits and quantum gates, manages to achieve better efficiency in solving the problem exactly, i.e. by finding each and only the attractors present in a general synchronous Boolean network. 

Given an ensemble of $n$ Boolean variables (each representing, for example, a gene of the GRN), one can always find $2^n$ possible states for the network. A set of rules (Boolean Functions, BFs) can then be used to determine their dynamics over time, i.e. how a given state of the system is continuously mapped to a new one (possibly itself again) at each time step. The collection of all $2^n$ possible system states and their dynamical evolution is called a state transition graph (STG). As mentioned in the introduction, the exact solution of such a graph scales exponentially. Thus, it can be proved that the problem of finding both static and cyclic attractors in BNs is NP-hard \cite{akutsu1998system, Akutsu2012, Mori2022}. This limits exhaustive simulations to small networks, e.g. using the BoolNet package in R, to $n \le 29$. For larger systems, heuristics are used that can guarantee to find the entire attractor set, but do not provide any additional information about the attractor basins. Such heuristics can be based on Boolean Satisfiability (SAT) \cite{dubrova2011sat, tamura2009detecting} or equivalent methods.

\begin{figure*}
\hspace*{-0.7cm}\includegraphics[width=1.1\linewidth,keepaspectratio]{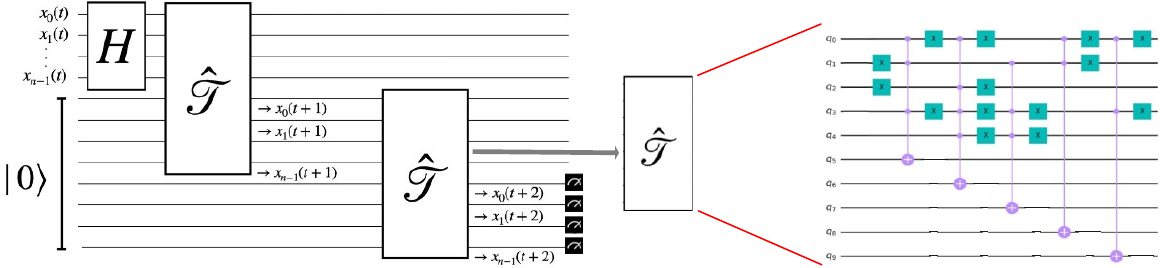}%
\caption{\label{fig:tev_example}: Schematic of a circuit implementing the time evolution dynamics of a given Boolean network. The scheme on the left produces a superposition of all attractors, and the final measurements will lead to finding one of them. On the right, an example of what a time evolution operator can look like for a given set of Boolean rules.}%
\end{figure*}

\begin{figure*}
\hspace*{-0.7cm}\includegraphics[width=1.1\linewidth,keepaspectratio]{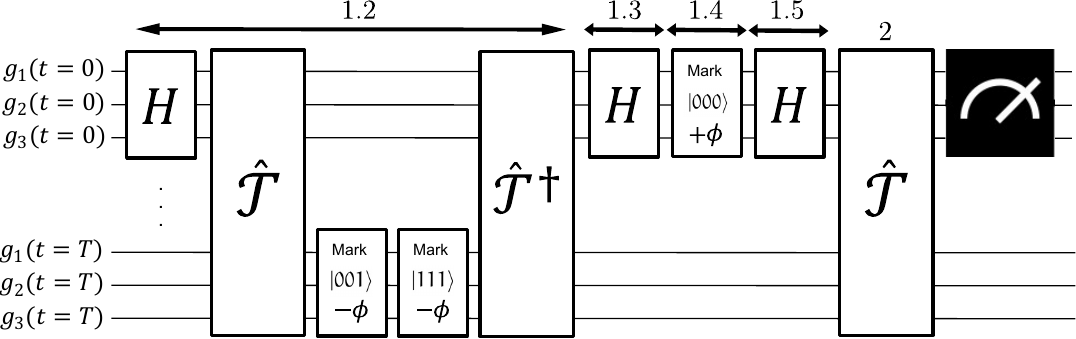}%
\caption{\label{fig:revGrover_example}: Schematic example of the basin suppression algorithm. In this example, drawn for 3 genes for better understanding, we show the suppression of two attractor basins of the network, namely the states $\ket{001}$ and $\ket{111}$. The numbered steps on top refer to the phases defined in Sec.\ref{sec:algorithm}, while the estimation of the basin size is omitted for better visualization of the attractors measurement routine. The final measurement (step 2), after the final time-evolution operator, will lead to the measurement of an attractor state not yet uncovered and suppressed.}%
\end{figure*}

The quantum approach is inspired by its classical counterpart, but exploits the possibilities offered by quantum superposition and quantum interference of qubit states. As we show in Fig.\ref{fig:revGrover_example}, it can be divided into two main \textit{parts}, the first being then further subdivided in 5 \textit{steps} (a detailed breakdown of the algorithm is provided later in Sec.\ref{sec:algorithm}). The first part consists of initialising the qubit system to the desired initial state. Such a state, in the first execution of the algorithm on the quantum device (from now on called \textit{runs} of the algorithm), is the superposition of all possible $2^n$ states that the system can be in. As often done in the literature, it can be obtained by applying a Hadamard gate to each qubit. Then, remembering that each possible state of the network represents an orthogonal dimension in the Hilbert space of all the qubits, in the second part, we apply the BFs to such a prepared state, thus evolving each state belonging to the superposition into each time-evolved state. In the first run, by evolving the initial Hadamard state a sufficient number of times, we will eventually recover a superposition of all and only the static attractors of the network and elements of the dynamical attractors. A small example is shown in Fig.\ref{fig:tev_example}. The measurement of such a state would now result in one of the attractors of the system (or a state belonging to a dynamical attractor). In their recent work, Weidner et al\cite{weidner2023leveraging} successfully implement such BFs on a quantum device and couple them with a Grover search algorithm to efficiently explore the basins of attraction of the attractors of a network, provided, however, that such attractors are known in advance. A more detailed description of such a methodology can be found instead in Weidner et al\cite{weidner2023method}. 

The next step is, therefore, to be able to efficiently measure all the different attractors of a given network, possibly avoiding getting the same result multiple times (e.g. by measuring an attractor of the network with a particularly large basin of attraction multiple times, as happens in classical methods), thus optimizing the efficiency of the algorithm. So, once the first attractor of the system has been measured, in order to measure the next attractor, we will again apply the two steps described above, this time defining a different strategy for initializing the qubit system at the beginning: In fact, we want to develop a state preparation strategy capable of preparing the initial state in the superposition of all the possible states of the system \textit{but} the states belonging to the basin of attraction of the first attractor found. In this way we can ensure that our next measurement after applying the BFs to the system will lead to uncovering a different attractor than the one initially found. To this end, we have developed a novel algorithm that manages to apply an optimized amplitude-suppression strategy to this problem inspired by the principles of the Grover algorithm. A sketch of such an algorithm is shown in Fig.\ref{fig:revGrover_example}, while the details of our development of this algorithm can be found in the Algorithm section below. The iterative application of this method to any Boolean network would allow a different attractor to be found each time the algorithm is run, ensuring that all the system's attractors are found within a number of runs equal to the number of attractors present in the network. 

Finally, we would like to point out that this method works perfectly not only for so-called static attractors (single states in which the system remains once reached) but also for identifying dynamic attractors of a network (set of states through which the system repeatedly cycles). In the latter case, our algorithm would correctly identify a state belonging to the set of dynamic attractors. Uncovering the whole cycle is a trivial task that can be done later with a classical implementation of the Boolean maps that generate the GNR.

\begin{figure*}
\hspace*{-0.7cm}\includegraphics[width=1.1\linewidth,keepaspectratio]{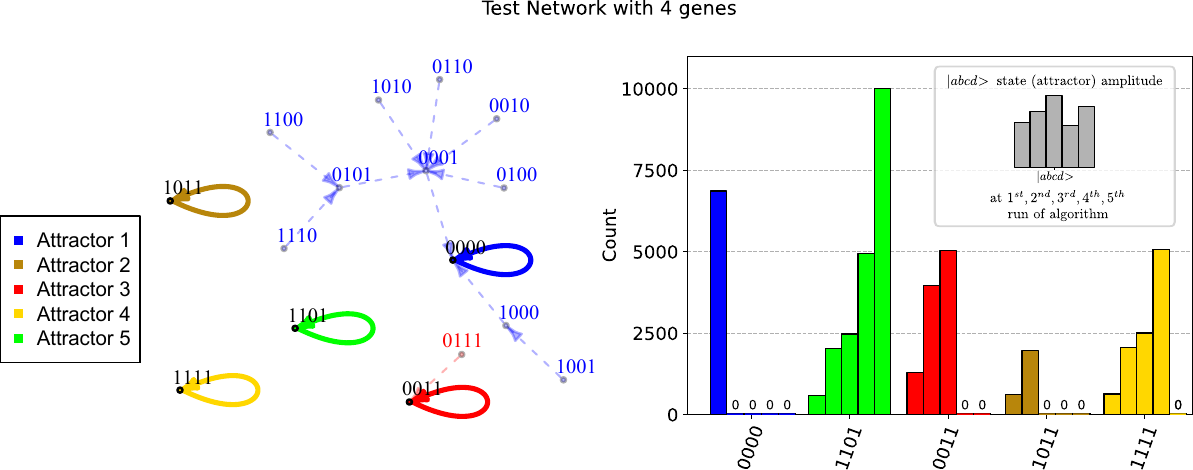}%
\caption{\label{fig:n4testnet_att}: On the left: Schematic representation of the Boolean network generated by 4 interacting genes designed ad hoc as a test case. The attractors are labeled with the order of the experimental measurement in our example. On the right: Count probability of measuring each attractor of the system after each run of the algorithm. The attractors found in the previous run are suppressed in the following runs. For clarity, the histogram highlights "0 counts" results.}%
\end{figure*}

\begin{figure*}
\hspace*{-0.7cm}\includegraphics[width=1.1\linewidth,keepaspectratio]{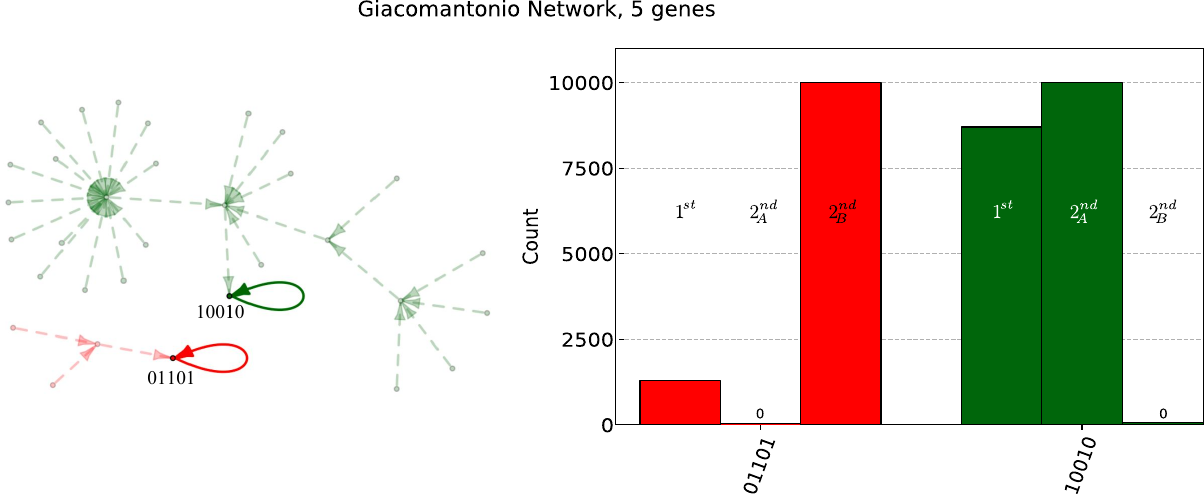}%
\caption{\label{fig:giaco_att}: On the left: Schematic representation of the Boolean network generated by 5 interacting genes described by Giacomantonio et al\cite{Giacomantonio2010}. Only the two attractors are marked for better readability. On the right: Count probability of measuring each attractor of the system. The three runs ($1^{st}, 2_A^{nd}$ and $2_B^{nd}$) describe respectively the measurement of the first attractor (probabilistic choice of one of the two) and the measurement of the second attractor depending on the result of the previous measurement. For clarity, the histogram highlights "0 counts" results.}%
\end{figure*}

\section{Examples}
We prove our results on two networks, one with four interacting genes created ad-hoc and a real one introduced by Giacomantonio et al\cite{Giacomantonio2010} with a real interaction map of five boolean agents describing mammalian cortical area development. In Figs.\ref{fig:n4testnet_att},\ref{fig:giaco_att} we show the networks and the probability histograms for each state to be measured after each run of the algorithm, removing at each run the previously found attractors. For the smaller network (Fig.\ref{fig:n4testnet_att}), we show a subsequent removal of attractors in an arbitrarily chosen sequence. For the Giacomantonio network (Fig.\ref{fig:giaco_att}), which features only two attractors, we show alternatively the removal of one or the other attractor from the superposition, showing how the algorithm exactly removes the respective basins of attraction regardless from their sizes (4 or 28 states). All results are performed on IBM Qiskit quantum device simulators. This allows us to demonstrate the determinism and exactness of the algorithm we present, and the low impact of shot noise on the expected results. In the next section we will present an analysis of the robustness of the algorithm against machine noise. Indeed, we can see in both cases how the impact of shot noise is not relevant to the scopes of the algorithm and how the algorithm exactly removes the required attractors from the state superposition up to leaving only the last to be found, with 100\% probability (10000 shots).

\section{Noisy runs}

\begin{figure*}
\hspace*{-0.7cm}\includegraphics[width=1.1\linewidth,keepaspectratio]{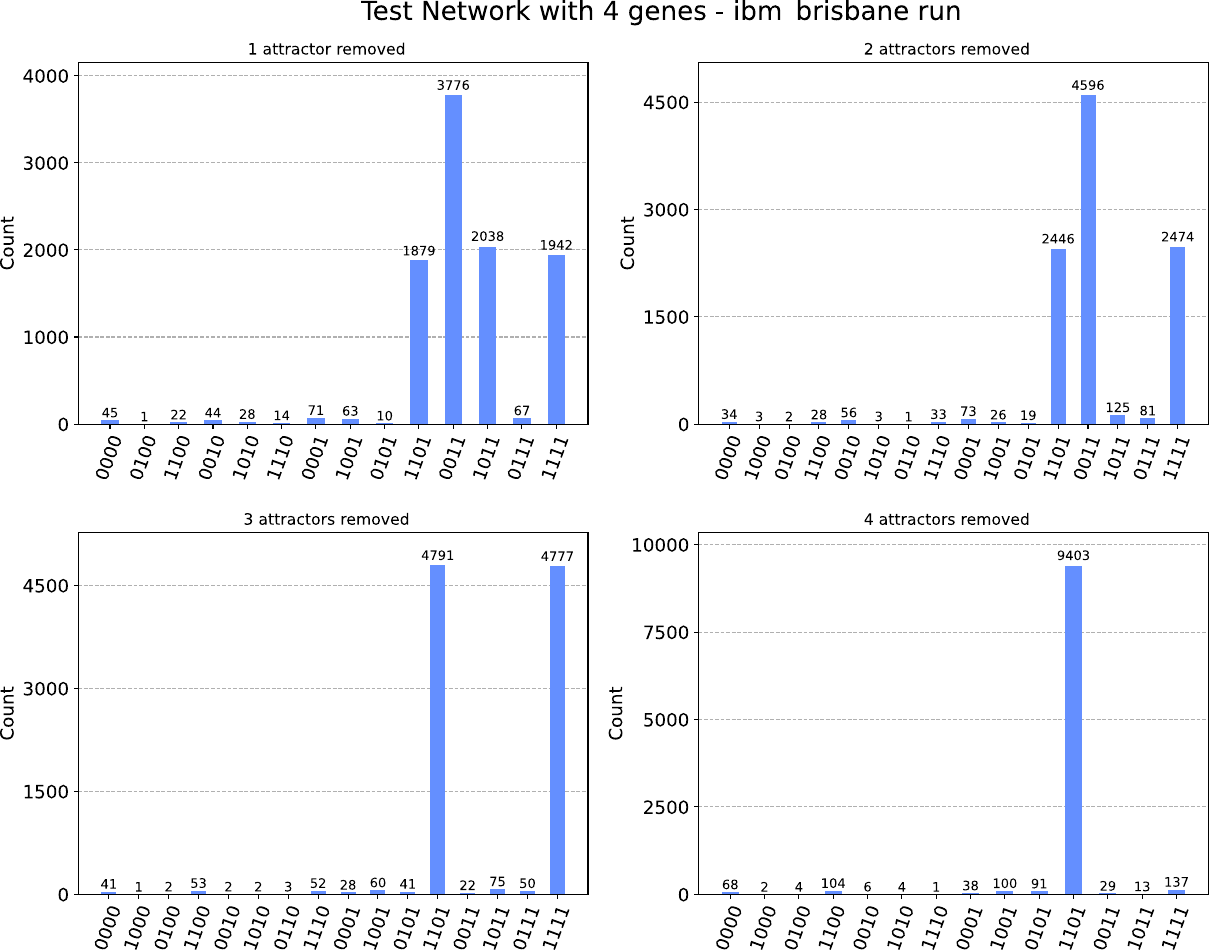}%
\caption{\label{fig:n4testnet_brisbane}: Subsequent runs of the deletion algorithm over the attractor basins of the 4-gene network introduced above. We prove our algorithm on a simulator that takes into account the noise profile extrapolated by the \textit{ibm\_brisbane} quantum device. In each run, we delete one more attractor basin (assumed to have been measured in the previous run) in the following order 0000 -> 1011 -> 0011 -> 1111. The last one to be measured will be 1101.}%
\end{figure*}

\begin{figure*}
\hspace*{-0.7cm}\includegraphics[width=1.1\linewidth,keepaspectratio]{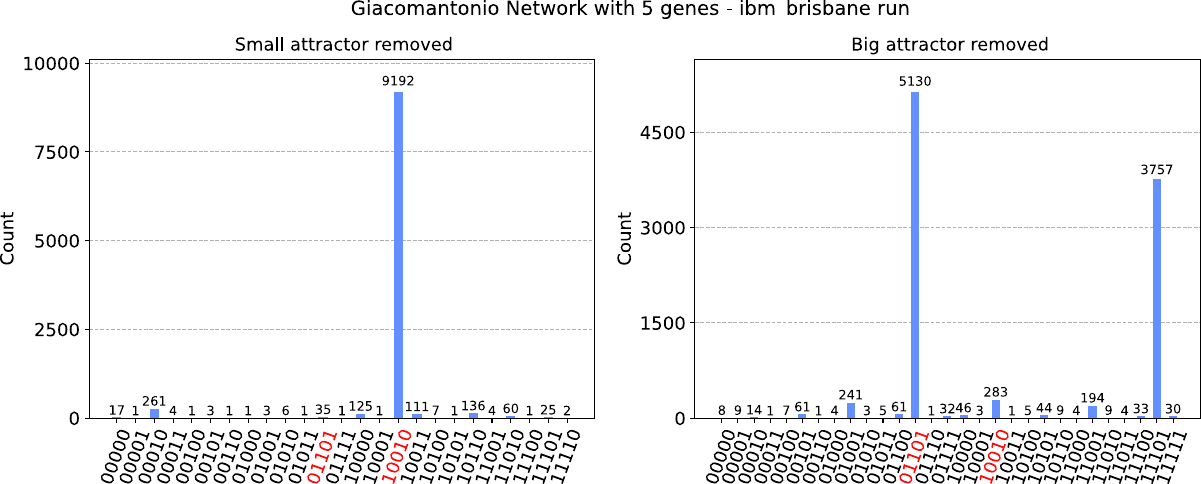}%
\caption{\label{fig:giaco_brisbane}: The deletion algorithm runs over the attractor basins of the Giacomantonio network introduced above. We prove our algorithm on a simulator, taking into account the noise profile extrapolated by the \textit{ibm\_brisbane} quantum device. In these two runs, we alternatively suppress the basin of either the small (left) or the large (right) attractor (highlighted in red in both plots), allowing the measurement of the other.}%
\end{figure*}

The qiskit package and IBMQ devices allow us to perform simulated runs of our quantum algorithm that mimic the error profile (the set of intrinsic probabilistic errors resulting from the imperfect NISQ nature of current quantum devices) of any machine currently available among the IBMQ quantum platforms. Since the choice of a particular platform among those currently available does not make a fundamental difference, we decided to use the noise profile of the \textit{ibm\_brisbane} platform. This allowed us to approximate the resistance of our algorithm to the noise generated by state-of-the-art quantum machines. The results are shown in figs.\ref{fig:n4testnet_brisbane},\ref{fig:giaco_brisbane}.

Fig.\ref{fig:n4testnet_brisbane} shows the same four runs that we have previously shown in the "non-noisy" simulations in Fig.\ref{fig:n4testnet_att}. Each histogram represents one run as we remove the attractors one by one, assuming they are removed in the same order as before. If we define the error probability as the probability of measuring a state from a run that does not belong to the set of attractors yet to be found, the error probabilities for each of these four runs are, in order, $3.65\%, 4.84\%, 4.32\%$ and $5.97\%$.

In fig. \ref{fig:giaco_brisbane} instead we find the results for the Giacomantonio network. In the two plots we can see the efficiency of the algorithm in removing both the small (left) and the large (right) attractor basins, thus allowing the measurement of the other attractor, assuming noisy conditions (\textit{ibm\_brisbane} noise simulator). The deletion of the small attractor gives rise to the measurement of the big attractor with $91.92\%$ probability, with a noisy "grass-like" distribution of wrong outcomes filling the missing $8.08\%$.

The task of removing the large attractor basin to allow the measurement of the small attractor gives a slightly different result. In fact, while we still find the $51.30\%$ probability of measuring the correct attractor and a "grass-like" distribution of small probabilities for most of the wrong results, we still find a $37.57\%$ probability of measuring a wrong result with a bit flip on the first qubit. However, such a situation can be easily overcome by introducing a hybrid quantum-classical routine, which consists of checking the nature of each algorithm result with a classical computational algorithm. In fact, checking the nature of a given state is a fast classical problem, which allows us to optimise the robustness of our algorithm against noisy results in a very efficient way.

Our results show how robust this algorithm is to the expected sources of noise from the state-of-the-art quantum devices in the IBMQ fleet. Although the technical implementation of Quantum Computing on a large and daily scale still requires efforts from the scientific community, this tool proves to be a useful and reliable technique to tackle a computational task that is widely used in bioinformatics, medicine and many other fields.

\section{Algorithm description \label{sec:algorithm}}
In the following, we will implement the BFs, provided in text files formatted by the BoolNet R package \cite{mussel2010boolnet}, on quantum circuits by making use of the ClassicalFunction compiler of the tweedledum package used by Qiskit \cite{cross2018ibm}. This way we can generate quantum circuits implementing $T$ state transitions of an $n$-component network using $(T+1)\cdot n$ qubits \cite{weidner2023leveraging}. As introduced in the previous section, the algorithm is made up of two parts, to be carried out consecutively on each run. 

\textit{The first part} consists of the development of a state initialisation routine that initialises the system in the superposition of all states belonging to the attraction basins of attractors that have not yet been found. On the first run of the algorithm, this part trivially consists of a layer of Hadamard gates alone. A seminal paper by Liu and Ouyang\cite{liu2013quantum} theoretically demonstrated a modification of Grover's algorithm that is able to delete a specific set of $M$ marked states from a uniform superposition over all $N=2^n$ basis states. Here, we have taken inspiration from their result to develop an algorithm (and its circuit implementation) adapted to fit our purpose. 

Our algorithm consists of five steps. The first step is to find out how many states we want to suppress from the superposition, after which we can write the operator $\hat{S}_t$ that performs the suppression of the selected states from a Hadamard state $\ket{\psi} = \hat{H}^{\otimes n} \ket{0}_n$ as follows:
\begin{equation*}
    \hat{S}_t\ket{\psi} = -\hat{H}^{\otimes n}\hat{I_0}\hat{H}^{\otimes n}(\hat{T}_t^\dagger\hat{I_c}\hat{T}_t)\ket{\psi}
\end{equation*}
where each operator represents one of the other four steps required for our state inizialization routine. We list and describe these 5 steps here and show our implementation on the quantum device.

\begin{itemize}
  \item Step 1.1: Estimate the number of states to be suppressed. Following the strategy used by Weidner et al\cite{weidner2023leveraging}, we run the quantum counting algorithm\cite{brassard1998quantum} once to obtain the number of states we want to erase from the total superposition (i.e. the number of states that belong to attractor basins already discovered). This parameter is used to estimate the angle $\phi$ used in steps 2 and 4.
  \item Step 1.2: Phase shift applied to all computational basis state not to be suppressed (not marked). This step is the most complex, as we want to \textit{not mark} a set of states (all states belonging to one or more known attractors) which are not in general known. To do so, after an initial layer of Hadamard gates, we first implement the operator $\hat{T}_t$, which evolves our system in time until the system converges to the superposition of the attractors. At this point, by applying the operator $\hat{I_c}$:
\begin{equation*}
    \hat{I_c} = \hat{I} + (e^{i\phi}-1)\sum_{i\neq\tau} \ket{i}\bra{i}
\end{equation*}
we apply the phase shift to all states that we do not want to suppress. To do this more efficiently, we instead mark with an opposite phase $-\phi$ the actual attractor(s) whose bases are to be suppressed, since a global shift of $+\phi$ to all states (corresponding to the identity operator) would then lead to the desired result. The angle $\phi$ will be specified later and $\tau$ is the basis state (or set of basis states) to be suppressed. The introduction of such tailored phase shifts, rather than the kickback method implemented in Grover's amplification algorithm, is the key to optimising the algorithm so that only one query is required for perfect suppression. It's derivation is shown in the work of Liu et al\cite{liu2013quantum}. The technical implementation of the phase shift gates, which apply the desired phase shift to the required states, is instead a generalisation of the algorithm developed by Fujiwara et al\cite{fujiwara2005}, where instead of a pi-shift we implement a more general $\phi$-shift. In the supplementary material we include a short description of the implementation of such a routine. Finally, we backpropagate the time evolution by applying $\hat{T}_t^\dagger$ to the system to mark all states that do not belong to the selected attractor(s). More details on the use of time evolution operators can be found in Weidner et al\cite{weidner2023method, weidner2023leveraging};
  \item Step 1.3: Layer of Hadamard gates applied to all qubits, $\hat{H}^{\otimes n}$; 
  \item Step 1.4: Conditional phase shift of $e^{-i\phi}$ applied to the $\ket{0}_n$ state and of $e^{i\pi}$ applied to all other states:
\begin{equation*}
    -\hat{I_0} = -\hat{I}-(e^{i\phi}-1)\ket{0}\bra{0}
\end{equation*}
where the $\phi$ has to match the one of Step 1. Adopting a similar strategy as in Step 1.2, we apply a global shift to the transformation leading us to only apply a phase shift of $e^{+i\phi}$ applied to the $\ket{0}_n$ state;
  \item Step 1.5: Layer of Hadamard gates applied to all qubits, $\hat{H}^{\otimes n}$.
\end{itemize}

A single application of this routine will ensure the exact suppression of a set of $M$ basis states from a uniform superposition of $N$ states, as long as the ratio $\beta = \frac{M}{N} < \frac{3}{4}$. More in general, the optimal number of iterations $J$ this subroutine requires to exactly suppress $M$ states is:
\begin{equation*}
    J = \left\lceil \frac{\pi}{2\pi-4\beta}-\frac{1}{2} \right\rceil,    
\end{equation*}
where $\lceil x \rceil$ represent the Ceiling function ceil($x$). Following this definition, we can now define the angle $\phi$ introduced above as:
\begin{equation*}
    \phi = 2\arcsin \left( \frac{ \sin \left( \frac{\pi}{4J+2} \right) }{\cos\left(\beta\right)}\right).
\end{equation*}
More theoretical insights about this procedure can be found in the paper of Liu et al\cite{liu2013quantum}.

After the suppression of the states we are not interested in measuring from the initial uniform superposition, in \textit{the second part} of the algorithm we simply apply the time evolution operator $\hat{T}_t$ again and then measure the system. The suppression performed in the first part of the algorithm ensures that we will now measure a different static or dynamic attractor than the ones we found before. A comprehensive sketch of the described method is presented in Fig.\ref{fig:revGrover_example}

\section{Conclusions}
This work introduces a novel method for quantum computation for the exact solution of the problem of attractor search in Boolean Networks. Such a problem is strongly motivated by research in various fields, finding applications in theoretical computation, physics and engineering research. Here we introduce the problem as a way to study the behaviour of Gene Regulatory Networks (GRNs). Indeed, Boolean networks are often used to model the regulatory relationships between genes, and attractors in these networks correspond to stable gene expression patterns that can represent different cellular states, such as cell types or conditions (e.g., healthy vs. diseased states). 

We find that our algorithm is theoretically able to detect all attractors of any Boolean Network, even the smallest ones, with 100\% accuracy, in a number of queries corresponding to the number of attractors present in the system itself. To the best of the authors' knowledge, no algorithm in the literature currently achieves such a result. Moreover, runs simulating the results of state-of-the-art noisy quantum machines show that this method is stable, on relatively small example systems, against the errors induced by the current NISQ nature of quantum devices. This finding is promising for the development of future more fault-tolerant quantum technologies. In addition, we would like to highlight how this method can be further improved (both in terms of efficiency and error tolerance) by integrating classical computational techniques, such as validating the true attractor nature of a state found as an attractor by the quantum algorithm, which can be performed in the range of seconds even on very large networks.

Finally, as an alternative to the gate-based circuit, we would like to mention the possibility of implementing such an algorithm on a quantum annealer. Such hardware is specialized for solving optimization problems in QUBO form, and there are already methods to implement the logical AND, OR, and NOT operations as constraints in this form. Thus, it may be possible to take advantage of the larger number of qubits available on quantum annealers (over 5000 at the time of writing) compared to gate-based quantum computers (maximum of 433 qubits on IBM processors). This may allow faster scaling towards the analysis of larger networks, although the overhead in the number of qubits required to encode the logical constraints must first be established.

~
\section{Code availability}
The code for performing the analyses shown in this work as well as generating the resulting visualizations is available at\\ \textcolor{red}{\texttt{\url{https://github.com/sysbio-bioinf/QuantumAttractorSearch}}}.

\section{Acknowledgements}
HAK acknowledges funding from the  German Federal Ministery of Education and Research (BMBF)  e:MED confirm (id 01ZX1708C). Furthermore, HAK acknowledges funding from the German Science Foundation (DFG, SFB 1074 (no.  217328187), and SFB 1506 (no. 450627322) and GRK HEIST (no. 288342734)).

\section{Author Contributions}
The authors MR and FMW contributed equally to this work.

\section{Competing Interests}
The authors declare no competing interests.

\bibliography{main.bib}
\bibliographystyle{apsrev4-2}
\end{document}


\title{Supplementary Material: Enhancing Attractor Search in Boolean Networks with Quantum Computation}

\author{Mirko Rossini$^{1,2,*}$, Felix M. Weidner$^{3,*}$, Joachim Ankerhold$^{1,2}$, Hans A. Kestler$^{3,\dagger}$}
\affiliation{$^1$Institute for Complex Quantum Systems, Ulm University, 89069 Ulm, Germany \\
$^2$Center for Integrated Quantum Science and Technology (IQST) Ulm-Stuttgart, Germany \\
$^3$Institute of Medical Systems Biology, Ulm University, 89069 Ulm, Germany}
\altaffiliation{$^*$Equal contribution, $^\dagger$To whom correspondence should be addressed: [hans.kestler@uni-ulm.de]}

\date{\today}
\maketitle

\section{Implementation of conditional phase shifts}
We utilize this Grover-based deletion of states and combine it with the gate-based implementation of Boolean logic circuits presented by Weidner et al. \cite{weidner2023leveraging} and described in detail in \cite{weidner2023method}.
For this purpose, it is required to implement conditional phase shift operators $\hat{\Phi}_j^n$ which applies a phase $e^{i\phi}$ the $j$-th basis state in an $n$ qubit system while leaving all others unchanged. Here, $j \in \{0,...,2^n - 1\}$, with $j$ corresponding to the basis states $0 = \ket{0...00}, 1 = \ket{0...01}, 2 = \ket{0...10}$ and so on.

Mathematically, such phase shift is described by the matrix:
\begin{equation}
    \hat{\Phi}_j^n =
    \begin{bmatrix}
        1 & & & &  &   \\
        & 1 &  & & \text{\huge0} &   \\
        & &  \ddots &  & &  \\
        &  & & e^{i\phi} \\
        & \text{\huge0}& & & \ddots &  \\
        & & & &  &  1 \\
    \end{bmatrix}
\end{equation}

\begin{figure}[!ht]
    \centering 
    \includegraphics[width=0.5\textwidth]{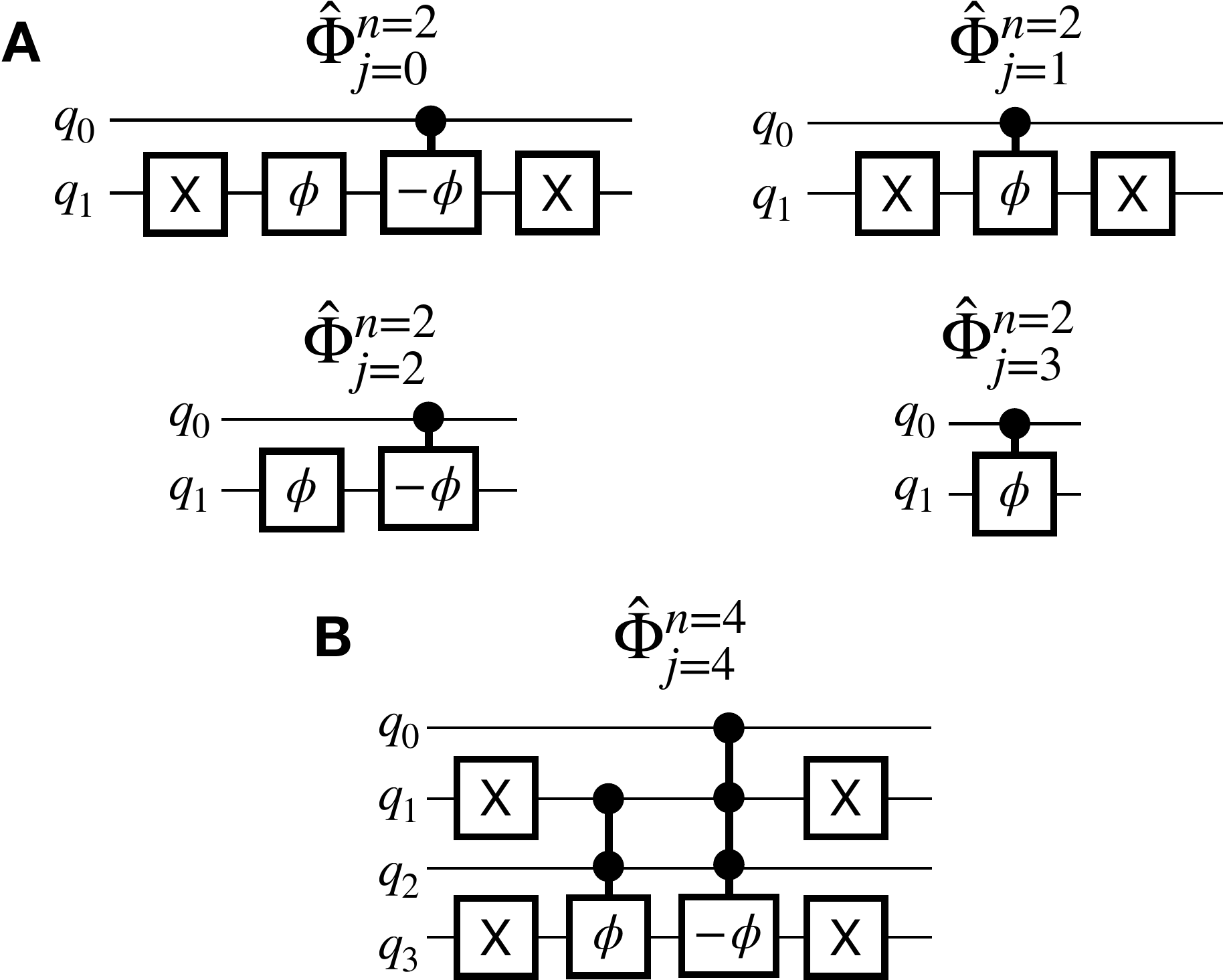}
    \caption{Quantum circuits for the conditional state shifters specified by the corresponding values of $n$ and $j$. 
        \label{fig:phaseshiftercircuits}}
\end{figure}

The phase $\phi$ for these operators is calculated based on the ratio of the number of marked bases states over the total amount of states $\tfrac{M}{N}$ in order to suppress completely the target states. The value of $M$ corresponds to the basin size of the already identified attractors and can be obtained by running a quantum counting circuit \cite{weidner2023leveraging, brassard1998quantum}.\\
A gate-based implementation of such phase shift operators is described by Fujiwara and Hasegawa \cite{fujiwara2005}.\\
These phase shift operators were specified by Fujiwara and Hasegawa for a base case of $n=2$ qubits as
\begin{equation}
    C_0^2 = \hat{\Phi}_2^2 = 
    \begin{bmatrix}
        1 & 0 & 0 & 0\\
        0 & 1 & 0 & 0\\
        0 & 0 & e^{i\phi} & 0\\
        0 & 0 & 0 & 1
    \end{bmatrix}, \quad
    C_1^2 = \hat{\Phi}_3^2 = 
    \begin{bmatrix}
        1 & 0 & 0 & 0\\
        0 & 1 & 0 & 0\\
        0 & 0 & 1 & 0\\
        0 & 0 & 0 & e^{i\phi}
    \end{bmatrix}.
\end{equation}
Based on this, phases can be applied to the remaining two basis states by construction of the operators $\hat{\Phi}_1^2 = N_c \hat{\Phi}_3^2 N_c$ and $\hat{\Phi}_0^2 = N_c \hat{\Phi}_2^2 N_c$, where $N_c$ indicates a layer of NOT gates to be applied to the qubits corresponding to the zeros of the basis state to shift. An example is provided in Fig.\ref{fig:phaseshiftercircuits}\\

\section{Scaling of the remaining state space after repeated iterations}

Up to $M/N=3/4$ of the remaining state space can be suppressed in $j=1$ iteration. Therefore, the fraction of basis states that can be suppressed in a space of size $N$ starting from a uniform superposition over all $N$ basis states in $j$ iterations of the Grover operator is $1-(\frac{1}{4})^j$.\\

The number of iterations $J$ that have to be performed is calculated depending on the ratio of marked states in the state space $M/N$ contained in the parameter $\beta$ as (\cite{liu2013quantum}): 
\begin{equation}
    \beta = \arcsin(\sqrt{M/N})
\end{equation}
\begin{equation}
    J = \Big\lceil \frac{\beta}{\pi - 2\beta} \Big\rceil.
\end{equation}
Based on this, the angle $\phi$ used in the phase shift operator is adapted to a value of
\begin{equation}
    \phi = -2 \cdot \arcsin\Big(\frac{\sin(\pi / (4J+2))}{\cos(\beta)}\Big)
\end{equation}
in every iteration of the Grover operator.


\bibliography{main.bib}
\bibliographystyle{apsrev4-2}
\clearpage